# Time-efficient, high-resolution 3T whole-brain relaxometry using Cartesian 3D MR-STAT with CSF suppression

Hongyan Liu[1], Edwin Versteeg[1], Miha Fuderer[1], Oscar van der Heide[1], Martin B. Schilder[1], Cornelis A. T. van den Berg[1], Alessandro Sbrizzi[1]

1. Computational Imaging Group, Department of Radiotheraphy, University Medical Center Utrecht, Utrecht, The Netherlands

**Corresponding authors:**

Hongyan Liu, University Medical Center Utrecht, Heidelberglaan 100, 3584CX Utrecht, the Netherlands; email: h.liu@umcutrecht.nl

Alessandro Sbrizzi, University Medical Center Utrecht, Heidelberglaan 100, 3584CX Utrecht, the Netherlands; email: a.sbrizzi@umcutrecht.nl

**Word count**: 2890 words

---

**Abstract (Word: 250)**

**Purpose:** Current 3D Magnetic Resonance Spin TomogrAphy in Time-domain (MR-STAT) protocols use transient-state, gradient-spoiled gradient-echo sequences that are prone to cerebrospinal fluid (CSF) pulsation artifacts when applied to the brain. This study aims at developing a 3D MR-STAT protocol for whole-brain relaxometry that overcomes the challenges posed by CSF-induced ghosting artifacts.

**Method:** We optimized the flip-angle train within the Cartesian 3D MR-STAT framework to achieve two objectives: (1) minimization of the noise level in the reconstructed quantitative maps, and (2) reduction of the CSF-to-white-matter signal ratio to suppress CSF signal and the associated pulsation artifacts. The optimized new sequence was tested on a gel/water-phantom to evaluate the accuracy of the quantitative maps, and on healthy volunteers to explore the effectiveness of the CSF artifact suppression and robustness of the new protocol.

**Results:** A new optimized sequence with both high parameter encoding capability and low CSF intensity was proposed and initially validated in the gel/water-phantom experiment. From in-vivo experiments with five volunteers, the proposed CSF-suppressed sequence shows no CSF ghosting artifacts and overall greatly improved image quality for all quantitative maps compared to the baseline sequence. Statistical analysis indicated low inter-subject and inter-scan variability for quantitative parameters in gray matter and white matter (1.6%–2.4% for T1 and 2.0%–4.6% for T2), demonstrating the robustness of the new sequence.

**Conclusion:** We presented a new 3D MR-STAT sequence with CSF suppression that effectively eliminates CSF pulsation artifacts. The new sequence ensures consistently high-quality, 1mm³ whole-brain relaxometry within a rapid 5.5-minute scan time.

**Abstract (Word: 250)**

**Purpose:** Current 3D Magnetic Resonance Spin TomogrAphy in Time-domain (MR-STAT) protocols use transient-state, gradient-spoiled gradient-echo sequences that are prone to cerebrospinal fluid (CSF) pulsation artifacts when applied to the brain. This study aims at developing a 3D MR-STAT protocol for whole-brain relaxometry that overcomes the challenges posed by CSF-induced ghosting artifacts.

**Method:** We optimized the flip-angle train within the Cartesian 3D MR-STAT framework to achieve two objectives: (1) minimization of the noise level in the reconstructed quantitative maps, and (2) reduction of the CSF-to-white-matter signal ratio to suppress CSF signal and the associated pulsation artifacts. The optimized new sequence was tested on a gel/water-phantom to evaluate the accuracy of the quantitative maps, and on healthy volunteers to explore the effectiveness of the CSF artifact suppression and robustness of the new protocol.

**Results:** A new optimized sequence with both high parameter encoding capability and low CSF intensity was proposed and initially validated in the gel/water-phantom experiment. From in-vivo experiments with five volunteers, the proposed CSF-suppressed sequence shows no CSF ghosting artifacts and overall greatly improved image quality for all quantitative maps compared to the baseline sequence. Statistical analysis indicated low inter-subject and inter-scan variability for quantitative parameters in gray matter and white matter (1.6%–2.4% for T1 and 2.0%–4.6% for T2), demonstrating the robustness of the new sequence.

**Conclusion:** We presented a new 3D MR-STAT sequence with CSF suppression that effectively eliminates CSF pulsation artifacts. The new sequence ensures consistently high-quality, 1mm$^3$ whole-brain relaxometry within a rapid 5.5-minute scan time.

## 1. Introduction

Three-dimensional multi-parametric quantitative MRI (qMRI) methods have been developed in order to acquire multiple volumetric tissue parameters with high signal-to-noise ratio (SNR) efficiency from one single scan. Examples of these 3D multi-parameter qMRI techniques include 3D MR Fingerprinting with optimized spiral trajectory [1], [2], 3D-quantification using an interleaved Look-Locker acquisition sequence with T2 preparation pulse (3D-QALAS) [3], [4] , quantitative transient-state imaging (QTI) [5], [6] and Magnetic Resonance Spin Tomography in Time-Domain (MR-STAT) [7], [8] .

MR-STAT is a model-based, transient-state multiparametric qMRI framework that allows for simultaneous estimation of $T_1$, $T_2$ and PD (proton density) from single short scans. In previous work [9], a 3D MR-STAT framework was developed to achieve higher SNR compared to 2D acquisitions and isotropic resolutions for 3D volumetric quantitative maps. Specifically, the proposed 3D MR-STAT sequence is a gradient-spoiled, gradient-echo (GRE) sequence using a repetitive slowly varying flip-angle train and a regular 3D Cartesian sampling trajectory. This 3D MR-STAT framework has been validated on healthy volunteers to acquire high-resolution knee and bilateral lower-leg quantitative maps. However, the previous protocol was not suitable for brain scans due to strong artifacts caused by cerebrospinal fluid (CSF) pulsation.

The CSF pulsatile flow has time-varying velocities which are driven by respiratory and cardiac cycles [10]; subsequently, the MR signal in CSF regions can be modulated by the varying flow velocity when using a flow-sensitive sequence, for example, an unbalanced GRE sequence. In 3D quantitative imaging with either steady-state or transient-state SSFP (Steady-State Free Precession) protocols [6], [9], [11], when neighboring phase-encoding k-space samples are acquired with significant CSF signal modulation, ghosting artifacts can be observed which originate from the CSF regions, e.g. cerebral ventricles, and affect surrounding tissues in the image domain [12]. These artifacts are more conspicuous when the CSF signal intensity is comparable or higher than that of the periventricular brain tissues, including deep grey matter (GM) structures such as the thalami, and the frontal white matter (WM).

A recently developed 3D $T_2$ mapping protocol [13], which used a steady-state, gradient-spoiled GRE sequence, incorporates a strategy to suppress the CSF signal intensity and the related ghosting artifacts. The idea is to use an optimized combination of the RF amplitudes and quadratic RF phases to achieve both good $T_2$ encoding for gray/white matter tissues and efficient suppression for the CSF signal intensity and the related ghosting artifacts. Inspired by this method, we apply a CSF suppression strategy in our multi-parametric, transient-states MR-STAT framework. Specifically, we optimize a time-varying flip-angle train for 3D MR-STAT to simultaneously achieve (1) signal suppression for CSF and (2) high SNR T1 and T2 of GM and WM. The resulting optimized 3D MR-STAT sequence enables whole-brain multi-parametric relaxometry in about 5.5 minutes at 1mm³ isotropic resolution without CSF ghosting artifacts and high quantitative image quality. The feasibility and robustness of the new protocol was validated with a water/gel phantom and five healthy volunteers at 3T MRI.

## 2. Methods
### 2.1 The baseline 3D MR-STAT sequence

MR-STAT enables the estimation of multi-parametric quantitative maps directly from transient-state time-domain data from one single scan. A typical 3D MR-STAT framework uses a fast gradient-echo, gradient spoiled sequence with a repetitive time-varying RF excitation flip-angle train [9]. A CAIPIRINHA-pattern Cartesian undersampling trajectory [14] can be applied to accelerate the acquisition. A waiting time $T_w$ and an initial adiabatic inversion pulse are typically used at the beginning of each repetition of the flip-angle train in order to achieve better T1 encoding. **Figure S1** shows a schematic example of the 3D Cartesian MR-STAT sequence.

In the previous work [9], a smoothly varying flip-angle train with four sine-square lobes was used in the 3D MR-STAT sequence. As illustrated in **Figure S1**, for each $kz$ location, the modulating flip-angle train covers the kx-ky plane for $N_k = 5$ times. This acquisition employs a 3D geometry with a matrix size of $N_x$ (in the readout direction) $\times N_y$ (in the fast phase-encoding direction) $\times N_z$ (in the slow phase-encoding direction), incorporating a CAIPIRINHA pattern[14] for an undersampling factor of 2 in both phase-encoding directions. Consequently, this leads to the length of the flip-angle train to be $5 * N_y/2$, and the number of flip-angle train repetitions to be $N_s * N_z/2$, where $N_s$ represents the slab oversampling factor in 3D acquisitions. A waiting time between repetitions of $T_w = 1.5s$ is introduced to improve $T_1$ encoding. **Figure 1(a)** shows the flip-angle train used as a baseline sequence in this work.

## 2.2 Proposed optimization of the CSF-suppressed RF phase and amplitude modulation

Within the MR-STAT framework, a methodology named BLock Analysis of a K-space-domain Jacobian (BLAKJac) [15] was developed to optimize the RF flip-angle train in order to achieve optimal SNR in the reconstructed parameter maps. In previous work, BLAKJac has been used to optimize 2D MR-STAT sequences, and it has also been shown that both the amplitude and the quadratic RF phase modulations can be optimized by BLAKJac to achieve better quantitative parameter encoding [16].

In this work, we extend the usage of BLAKJac to the 3D MR-STAT framework with multiple optimization goals. Specifically, we optimize both amplitude and phase of a time-varying flip-angle train for a 3D MR-STAT acquisition such that (1) the estimated noise spectrum for the reconstructed quantitative maps is low and (2) the CSF signal intensity is much lower than the neighboring brain tissues.

Let $\alpha(n)$ be the RF amplitude of the flip-angle train, and $\phi''(n)$ be the second-order RF phase derivative [16], respectively, with $n = 1,2,\dots,N_y/2 * 5$. The original Cramer-Rao bound (CRB) -based BLAKJac objective function can be defined as $F_o(\alpha,\phi'')$, which is a weighted sum of the noise level estimation for specific $(T_1, T_2)$ combinations. In order to suppress the CSF ghosting artifacts, we add an additional term to the original BLAKJac objective function $F_o(\alpha,\phi'')$, shown as below

$$F(\alpha,\phi'') = F_o(\alpha,\phi'') + \lambda \frac{\|S_{CSF}(\alpha,\phi'')\|_2^2}{\|S_{WM}(\alpha,\phi'')\|_2^2}. \tag{1}$$

Here, $S_{CSF}(\alpha,\phi'')$ and $S_{WM}(\alpha,\phi'')$ are the time-varying signals as computed by the Extended Phase Graph (EPG) model [17] for different tissue parameter values, namely $(T_1, T_2) = (4s, 2s)$ for $S_{CSF}$ and $(T_1, T_2) = (0.9s, 0.05s)$ for $S_{WM}$. The new objective function $F(\alpha,\phi'')$ is optimized such that the signal norm for CSF is relatively small comparing to signal norm for white matter, and therefore the artifacts originating from CSF flow could be reduced significantly. The weighting parameter $\lambda$ balances the CSF suppression term and the original CRB-based BLAKJac term $F_o(\alpha,\phi'')$.

## 2.3 MRI experiments
### 2.3.1 Sequence design and optimization

The following sequence parameters are used for both the baseline sequence and the proposed CSF-suppressed sequence to be optimized by BLAKJac: matrix size $N_x \times N_y \times N_z = 224 \times 224 \times 134$, resolution = 1mm³, TR = 7ms, slice oversampling factor $N_s = 1.28$, CAIPIRINHA undersampling factor = $2 \times 2$. Unlike the baseline sequence where a waiting time of $T_w = 1.5s$ is required for better $T_1$ encoding, a non-stopping pattern with $T_w = 0s$ is used for the new sequence optimization [18], therefore resulting in a shorter acquisition time.

Seven different combinations of $(T_1, T_2)$ pairs are used to compute the CRB-based objective term $F_o(\alpha, \phi'')$, with $T_1$ ranging from 0.3s to 2.3s and $T_2$ ranging from 0.03s to 0.2s. The new optimized sequence with RF amplitude $\alpha(n)$ and phase $\phi''(n)$ is obtained by optimizing equation (1) using the Nelder–Mead algorithm [19]. $\alpha(n)$ and $\phi''(n)$ are constrained to be cubic-spline functions with 5 control points in order to reduce the search space of the optimization problem [19]. The optimal $\alpha(n)$ and $\phi''(n)$ are selected from 10 repetitions of the optimization process with different randomized initial values to avoid issues of local minima in the optimization process. The weighing parameter $\lambda = 10$ is selected such that the optimized new sequence has a similar noise level evaluated by $F_o(\alpha, \phi'')$ comparing to the baseline sequence, and a $\|S_{CSF}\|/\|S_{WM}\|$ signal ratio as small as possible for CSF signal suppression.

### 2.3.2 Phantom and in-vivo experiments

The 3D MR-STAT sequences, the baseline sequence as well as the new optimized sequence (**Figure 1**), were implemented on a 3T MR scanner (Philips, The Netherlands). Both sequences were used for 3D scans of a phantom and in-vivo brains using a 32-channel head coil. For the baseline sequence, the total scan time was 7 min 46 sec for the whole-brain MR-STAT acquisition. For the proposed optimized sequence, the zero-waiting time $T_w$ led to a reduced scan time of 5 min 37 sec.

A separate low-resolution fast reference scan was used for reconstructing the receive coil sensitivity maps by the ESPIRiT method [20]. The $B_1^+$ inhomogeneities were estimated using a separate fast multislice $B_1^+$ [21] sequence with 3.5mm$^3$ isotropic resolution.

For the phantom experiments, nine gel-vials with different relaxation properties (TO5, Eurospin II test system, Scotland) and one tap-water vial to mimic CSF were scanned using both 3D MR-STAT sequences. For validation, gold standard $T_1$ values were acquired using inversion-recovery spin-echo sequences with fifteen inversion times between 50 ms and 3800 ms, and gold standard $T_2$ values were acquired using single-echo spin-echo sequences with fifteen echo times between 10 ms and 1600 ms.

Five volunteers (three male, two female, age $25 - 35$) were scanned using both the baseline and the new optimized sequence to assess the CSF suppression performance of the proposed new sequence. To further assess the in-vivo repeatability of the proposed new sequence, one volunteer was repetitively scanned four times in three separate scan sessions. Written informed consent was obtained from all volunteers in accordance with the local institutional Review Board. To compute and compare the relaxometric statistics, GM and WM were automatically segmented using the FSL toolbox [22], [23] based on $R_1(1/T_1)$ maps.

### 2.3.3 Reconstruction

A two-step reconstruction strategy as described in previous work [9] was used for the 3D MR-STAT reconstruction: we first decoupled the undersampled 3D data into into 2D 'fully-sampled' MR-STAT data by employing a SENSE reconstruction [24]; subsequently, we run slice-by-slice 2D MR-STAT reconstructions independently along the slow phase encoding direction (feet-head in this case). For the SENSE reconstruction (first sub-step) of the in-vivo data, an L1 wavelet regularization term was used [25]. For the second sub-step, the split slice-by-slice 2D MR-STAT reconstructions were performed on a desktop PC by an accelerated MR-STAT reconstruction algorithm [26] incorporating both a surrogate model for fast signal computations and an alternating direction method of multipliers (ADMM). The imperfect slab-profile and $B_1^+$ map were taken into account in the physics model as scaling factors for the effective flip angle experienced in each voxel.

## 3. Results

### 3.1 The optimized new sequence

**Figure 1** shows the flip-angle trains of (a) the baseline sequence [9] and (b) the optimized sequence with CSF suppression. The simulated MR signals, $S_{CSF}$ and $S_{WM}$ for the two sequences are plotted in the second row. The proposed sequence has a much lower CSF-to-WM signal ratio (0.386) compared to the baseline sequence (1.106), indicating an efficient suppression of the CSF.

### 3.2 Phantom experiment

**Figure 2** shows the water/gel-phantom experimental results. In **Figure 2(a)**, the SENSE reconstruction results of the central slice after reconstruction sub-step 1 are shown for both the baseline sequence and the proposed new sequence. The water tube (purple circle) is selected to mimic CSF and the gel tube no. 4 (orange circle, ground truth $T_1 = 0.646s$ and $T_2 = 0.078s$) is selected to mimic white matter tissue. Relatively low CSF-to-WM ratio can be observed in all five k-spaces for the proposed new sequence, whereas relatively high CSF signal can be observed in the 1$^{st}$, 4$^{th}$ and 5$^{th}$ k-spaces for the baseline sequence. The averaged CSF-to-WM ratio over five k-spaces for the proposed and baseline sequence are 0.32 and 4.0, respectively. These results demonstrate the successful signal suppression for tissues/fluids with high $T_1$ and $T_2$ values.

**Figure 2(b) and 2(c)** shows a comparison of the reconstructed $T_1$ and $T_2$ results from both the baseline sequence and the proposed sequence with CSF suppression. **Figure 2(c)** summarizes the reconstructed $T_1$ and $T_2$ values for each gel tubes. Mean gel-tube $T_1$ and $T_2$ values computed from both sequences show good agreements with the gold standard results. The standard deviation (std) values are similar for the two sequences, with a slightly higher (11.9%) $T_1$ relative std observed from the proposed sequence, and a slightly higher (28.5%) $T_2$ relative std observed for the baseline sequence. When considering the sequence efficiency defined as the $T_{1,2}$ SNR divided by the square root of the scan time per slice [27] , we find that the $T_1$ sequence efficiency is 18.8 for the baseline sequence and 19.8 (4.9% higher) for the proposed sequence, and the $T_2$ sequence efficiency is 9.7 for the baseline sequence and 14.7 (50.9% higher) for the proposed new sequence.

### 3.3 In-vivo experiments

**Figure 3** shows in-vivo 3D MR-STAT results from both the baseline and the proposed CSF-suppressed sequence. Representative slices of quantitative maps in all three mutually orthogonal orientations are shown for one volunteer. In the zoomed-in plot in Figure 3(b), it can be seen that CSF ghosting artifacts corrupt the baseline sequence results, and the CSF ghosting artifacts are mainly along the slow phase-encoding direction (feet-head), contaminating the image quality in surrounding tissues. The proposed CSF-suppressed sequence shows overall greatly improved image quality for all quantitative maps. Similar figures for other volunteers are included in the Supplementary material (Figure S2 to S4) and confirm these findings.

**Figure 4** shows the in-vivo results for five healthy volunteers using the proposed sequence with CSF suppression. Figure 4(a) shows one representative transverse slice for each of the five volunteers, and Figure 4(b) shows the whole-brain $T_1$ and $T_2$ distributions for the gray and white matter of each volunteer. The mean reconstructed $T_1$ and $T_2$ values are summarized in Figure 4(c). The reconstructed $T_1$ was $912 \pm 21.7$ ms (2.4% inter-subject variation) in white matter and $1385 \pm 33.6$ ms (2.4% inter-subject variation) in gray matter. The reconstructed $T_2$ was $35.0 \pm 1.6$ ms (4.6% inter-subject variation) in white matter and $49.7 \pm 1.8$ ms (3.6% inter-subject variation) in gray matter. Overall, the observed $T_1$ and $T_2$ values fall into the range of values reported in previous literature[18], [28]–[30].

**Figure 5** shows repeatability results for one volunteer using the proposed sequence. Figure 5(a) shows one transverse slice for four repeating scans, and Figure 5(b) and 5(c) summarizes the reconstructed $T_1$ and $T_2$ values for each of the four scans. The inter-scan variations are overall 35% lower compared to the inter-subject variations: the reconstructed $T_1$ was $965 \pm 17.7$ ms (1.8% inter-scan variation) in white matter and $1436 \pm 22.3$ ms (1.6% inter-scan variation) in gray matter; the reconstructed $T_2$ was $36.0 \pm 1.0$ ms (2.8% inter-scan variation) in white matter and $50.5 \pm 1.0$ ms (2.0% inter-scan variation) in gray matter.

The in-vivo results demonstrate the robustness of our proposed new sequence. The 3D volumetric reconstruction results for all volunteers are available online (https://gitlab.com/HongyanLiu/3D-MR-STAT-brain).

## 4. Discussion

In this work, we incorporated a CSF suppression strategy into the 3D MR-STAT framework to mitigate the CSF-induced pulsation artifacts. This was achieved by optimizing a cyclic flip-angle train with a mixed objective function, in order to acquire both high SNR $T_1$, $T_2$ and PD quantitative maps and low CSF signal. Phantom experiments show that the new optimized sequence preserves high accuracy of the reconstructed quantitative maps, and similar noise level compared to the baseline sequence. The higher $T_2$ sequence efficiency for the proposed sequence benefits from the application of quadratic RF phase which allows for better $T_2$ encoding [31], [32].

The proposed 5.5-minute CSF-suppressed sequence for whole-brain 1mm³ isotropic relaxometry was tested on 5 healthy volunteers. Compared to the baseline sequence, the overall image quality was greatly improved due to the proposed CSF artifact suppression strategy. We showed that the optimized cyclic sequence with locally quadratic RF phase modulation can achieve in-vivo quantitative maps with both high SNR and sufficient CSF artifact suppression, which leads to good image quality of the reconstructed quantitative maps.

Our CSF-suppression strategy works efficiently for quantitative neural imaging because it relies on the relaxation time differences between CSF and other brain tissues. For in-vivo brain, the GM and WM have relatively short $T_1$ (< 2 sec) and $T_2$(< 0.1 sec) values, but CSF usually has long $T_1$ (typically > 3 sec) and $T_2$ (typically > 0.2 sec). This CSF-suppression strategy based on relaxation time differences may be extended for other applications, for example, achieving suppression of blood for cardiovascular imaging [33], and these possibilities will be explored and investigated in the future.

Beside the proposed CSF signal intensity suppression strategy, there are other methods being developed to suppress pulsatile flow artifacts. For example, flow compensation techniques by adjusting imaging gradients can be applied to reduce the signal's phase contribution caused by flow [34]–[36]. Another technique developed recently designs a randomized order of phase encoding to eliminate the repeating artifacts induced by semi-periodic physiological signal fluctuations [37]. Future work may also include the comparison or combined application of these methods with the proposed CSF-suppression method.

In this work, the proposed 3D MR-STAT sequence for whole-brain relaxometry has been validated on healthy volunteers. Numerous studies have demonstrated that alterations in quantitative tissue parameters are linked to neurological diseases, including multiple sclerosis (MS), epilepsy and Parkinson's disease [38]–[42]. We are confident that our protocol holds significant promise for clinical research, and advancing the application of our technique will be a primary focus of our future work.

## 5. Conclusion

We demonstrated a new Cartesian 3D MR-STAT sequence with inherent CSF suppression that mitigates the CSF pulsation artifacts and consistently obtained high quality, 1mm$^3$ whole-brain relaxometry within 5.5-minute scan time.

**FIGURES**

**Figure 1.** The flip-angle train (Amplitude + Phase) and the simulated MR signal responses of (a) the baseline sequence and (b) the proposed CSF-suppressed sequence. The corresponding simulated signals for CSF and WM tissues are also shown.

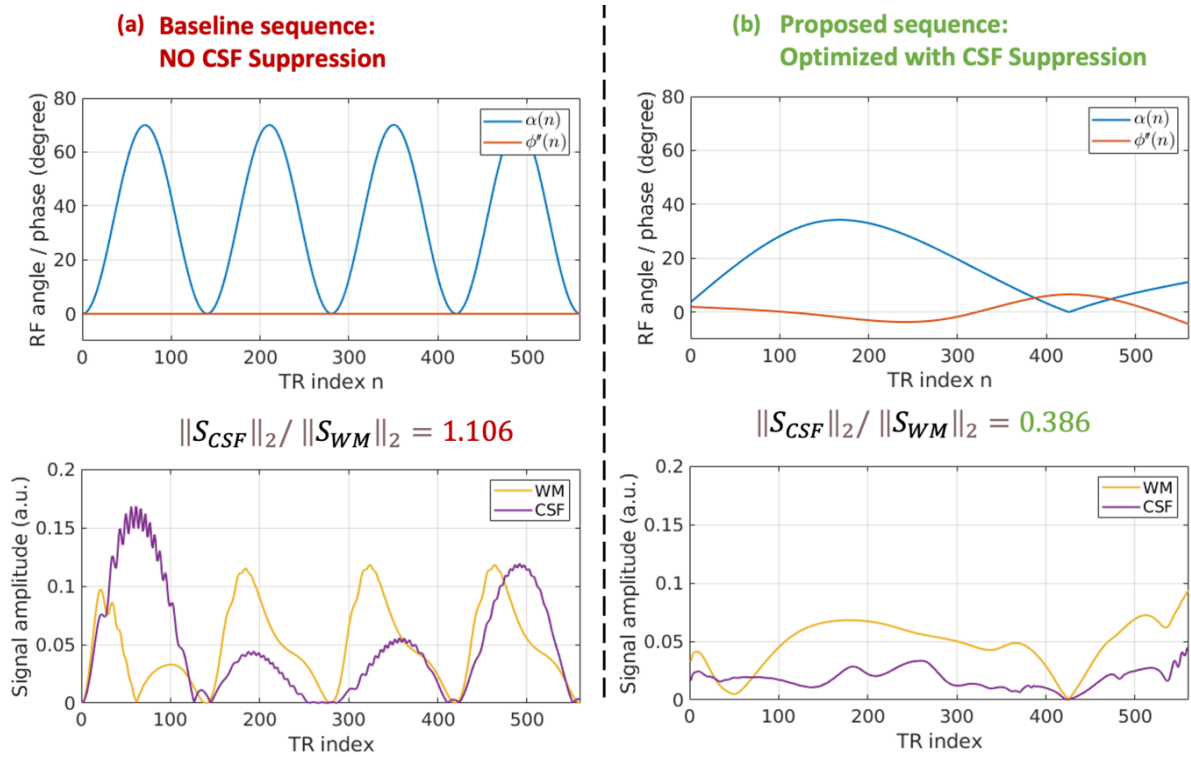

**Figure 2**. 3D MR-STAT phantom results from the baseline sequence and the new optimized sequence. (a) The SENSE reconstruction results after reconstruction sub-step 1. Image-domain results of the central slice for each of the five k-spaces are shown here for both sequences. (b) Quantitative $T_1$ and $T_2$ maps from both sequences after reconstruction sub-step 2. (c) Mean and standard deviations of the $T_1$ and $T_2$ values in 9 gel tubes vs the gold standard results.

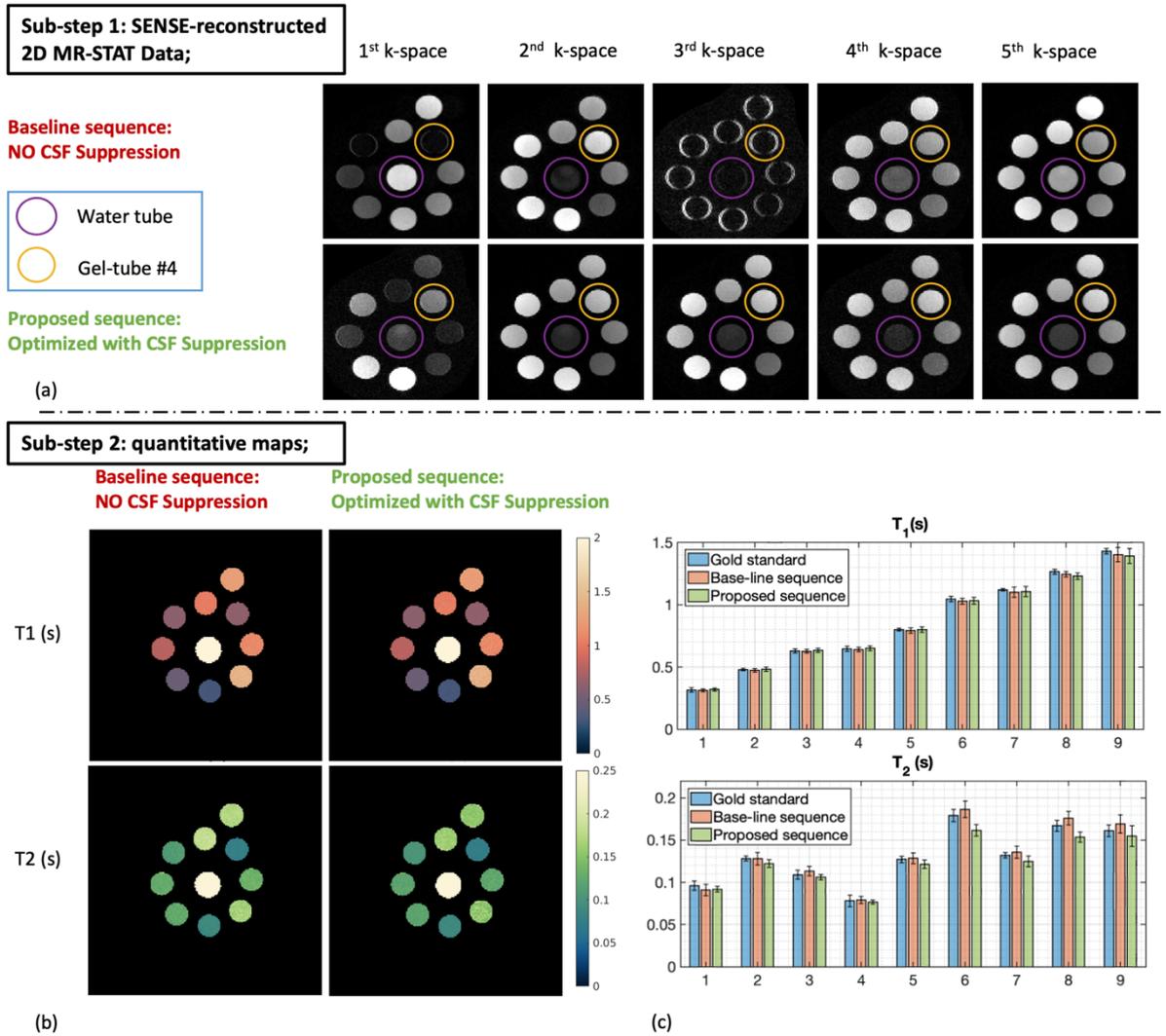

**Figure 3.** In-vivo 3D MR-STAT results using the baseline and the proposed CSF-suppressed flip-angle trains. Representative transverse, coronal and sagittal slices of quantitative maps are shown for one healthy volunteer. Inside the zoom-in frames, CSF ghosting artifacts can be seen for the baseline sequence reconstructions in the tissues that surround the lateral ventricle, but are removed in the proposed optimized sequence with CSF suppression. AP (anterior-posterior): frequency-encoding direction; LR (left-right): fast phase-encoding direction; FH (feet-head): slow phase-encoding direction.

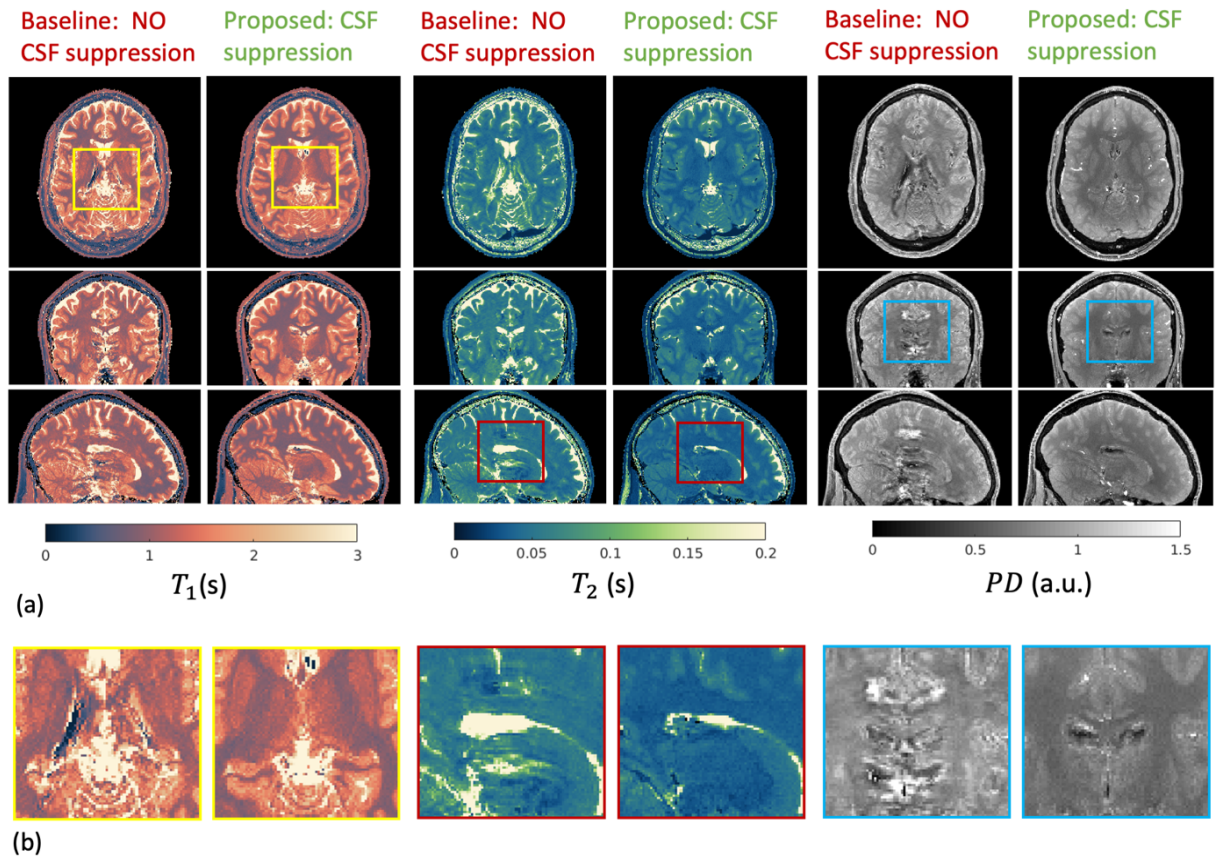

**Figure 4.** In-vivo results for five healthy volunteers using the proposed 3D MR-STAT sequence with CSF suppression. (a) One representative transverse slice for different volunteers; (b) The distribution of $T_1$ and $T_2$ values across different volunteers; the solid lines are the mean values, and the light-colored area show the standard deviation; (c) Reconstructed $T_1$ and $T_2$ values; literature values are from previous reference[18], [28]–[30].

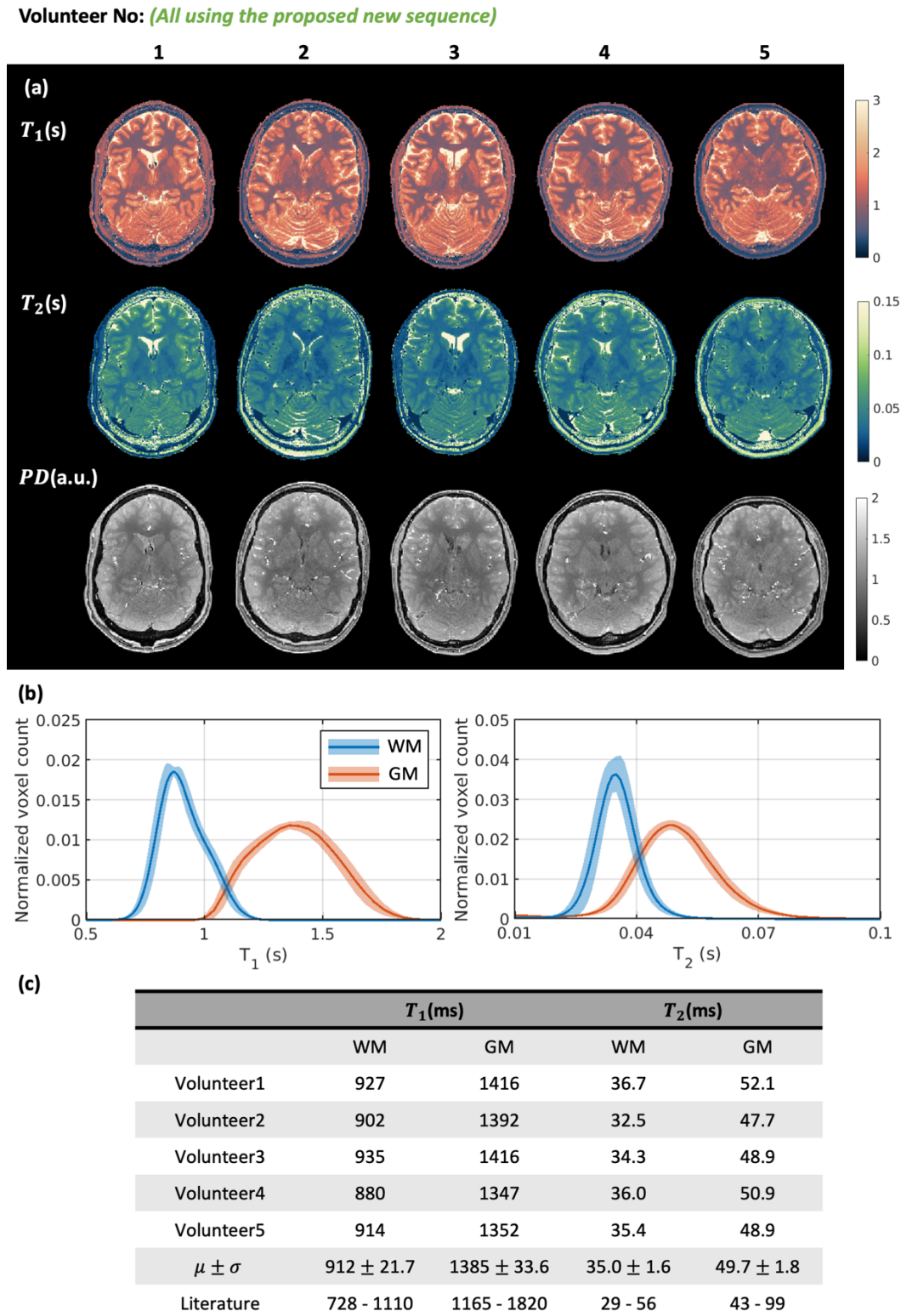

Volunteer No: *(All using the proposed new sequence)*

| | $T_1$(ms) | | $T_2$(ms) | |
|---|---|---|---|---|
| | WM | GM | WM | GM |
| Volunteer1 | 927 | 1416 | 36.7 | 52.1 |
| Volunteer2 | 902 | 1392 | 32.5 | 47.7 |
| Volunteer3 | 935 | 1416 | 34.3 | 48.9 |
| Volunteer4 | 880 | 1347 | 36.0 | 50.9 |
| Volunteer5 | 914 | 1352 | 35.4 | 48.9 |
| $\mu \pm \sigma$ | $912 \pm 21.7$ | $1385 \pm 33.6$ | $35.0 \pm 1.6$ | $49.7 \pm 1.8$ |
| Literature | 728 - 1110 | 1165 - 1820 | 29 - 56 | 43 - 99 |

**Figure 5.** Repeatability experiment results using the proposed 3D MR-STAT sequence with CSF suppression. One volunteer was repetitively scanned for 4 times. The volunteer was repositioned after scan 1 and scan 3. (a) One representative transverse slice for different scans; (b) The distribution of $T_1$ and $T_2$ values across different scans; the solid lines are the mean values, and the light-colored area show the standard deviation; (c) Reconstructed $T_1$ and $T_2$ values.

**Scan No:** *(All using the proposed new sequence)*

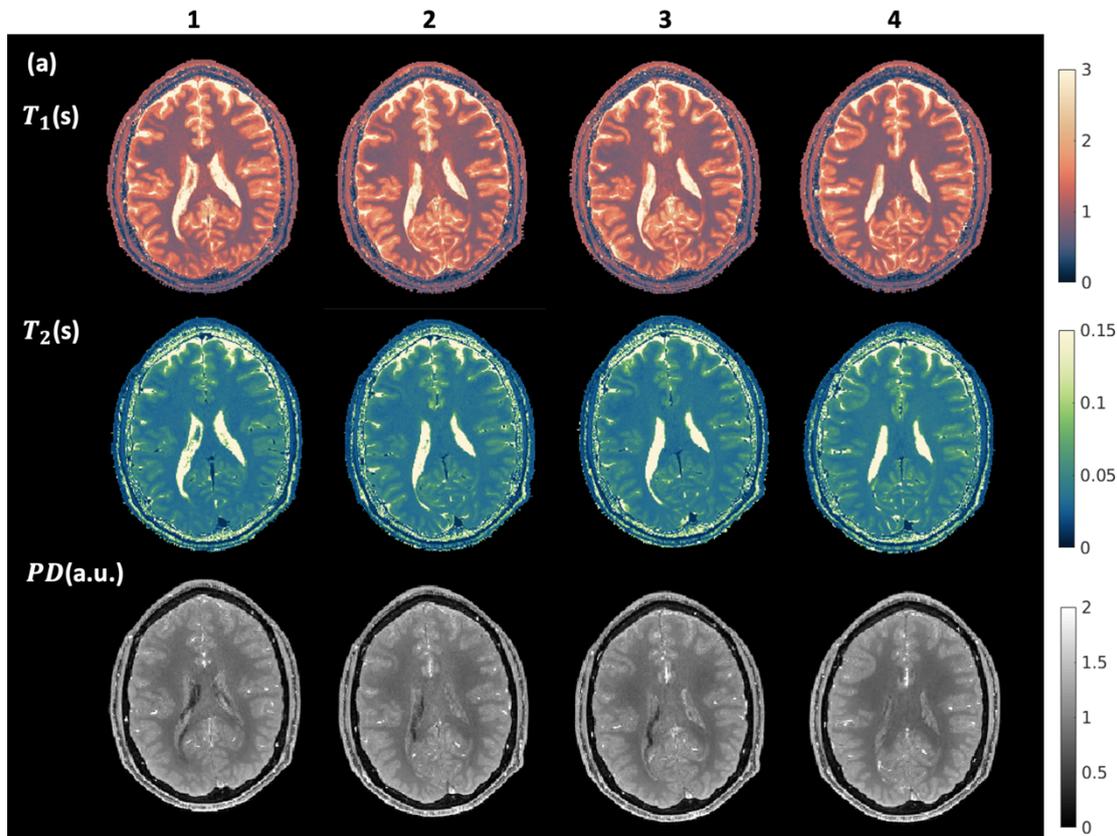

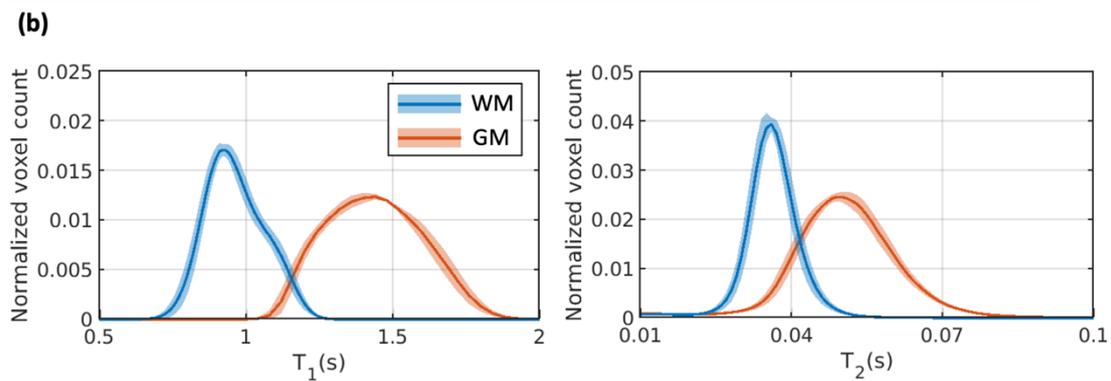

**(c)**

| | | $T_1$(ms) | | $T_2$(ms) | |
|---|---|---|---|---|---|
| | | WM | GM | WM | GM |
| Session1 | Scan1 | 975 | 1445 | 35.0 | 49.8 |
| Session2 | Scan2 | 974 | 1446 | 35.9 | 50.1 |
| | Scan3 | 971 | 1451 | 35.8 | 50.1 |
| Session3 | Scan4 | 938 | 1403 | 37.4 | 52.0 |
| $\mu \pm \sigma$ | | $965 \pm 17.7$ | $1436 \pm 22.3$ | $36.0 \pm 1.0$ | $50.5 \pm 1.0$ |

**Supplementary material**

**Figure S1.** Schematic example of a 3D Cartesian MR-STAT sequence with a CAIPIRINHA sampling factor of 2 x 2 in ky-kz phase encoding plane. The yellow trajectory shows the sampling order. An initial adiabatic inversion pulse is used at the beginning of each repetition of the flip-angle train, and a waiting time $N_w$ is used between repetitions.

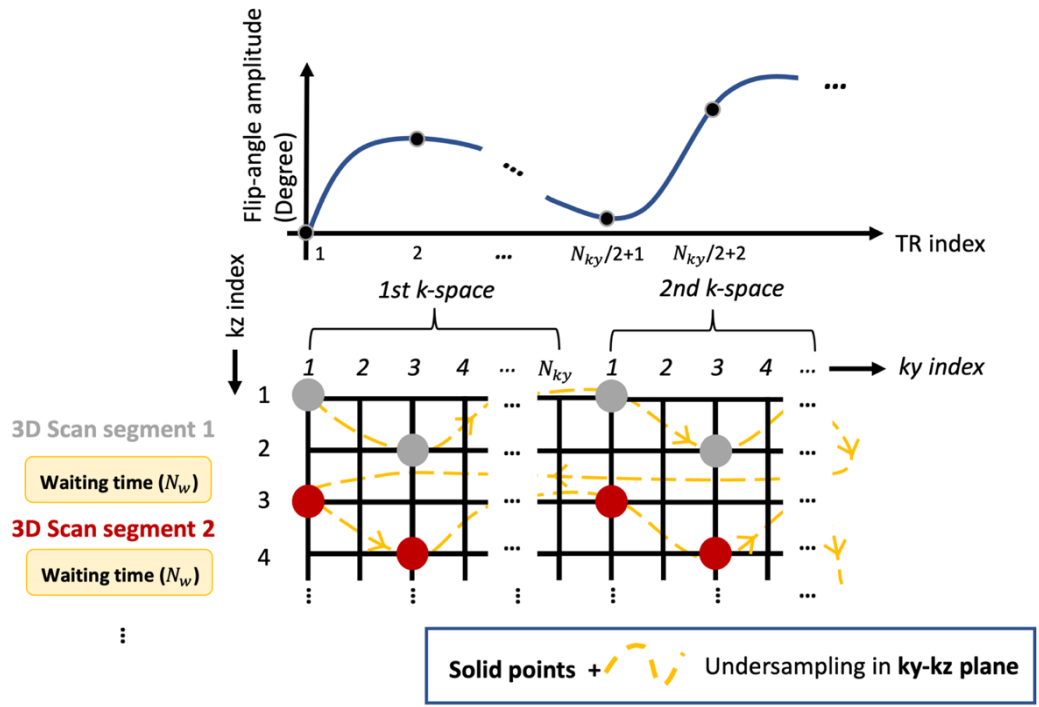

**Figure S2.** In-vivo 3D MR-STAT results using the baseline and the new optimized flip-angle trains. Similar as Figure 3 but for volunteer 2.

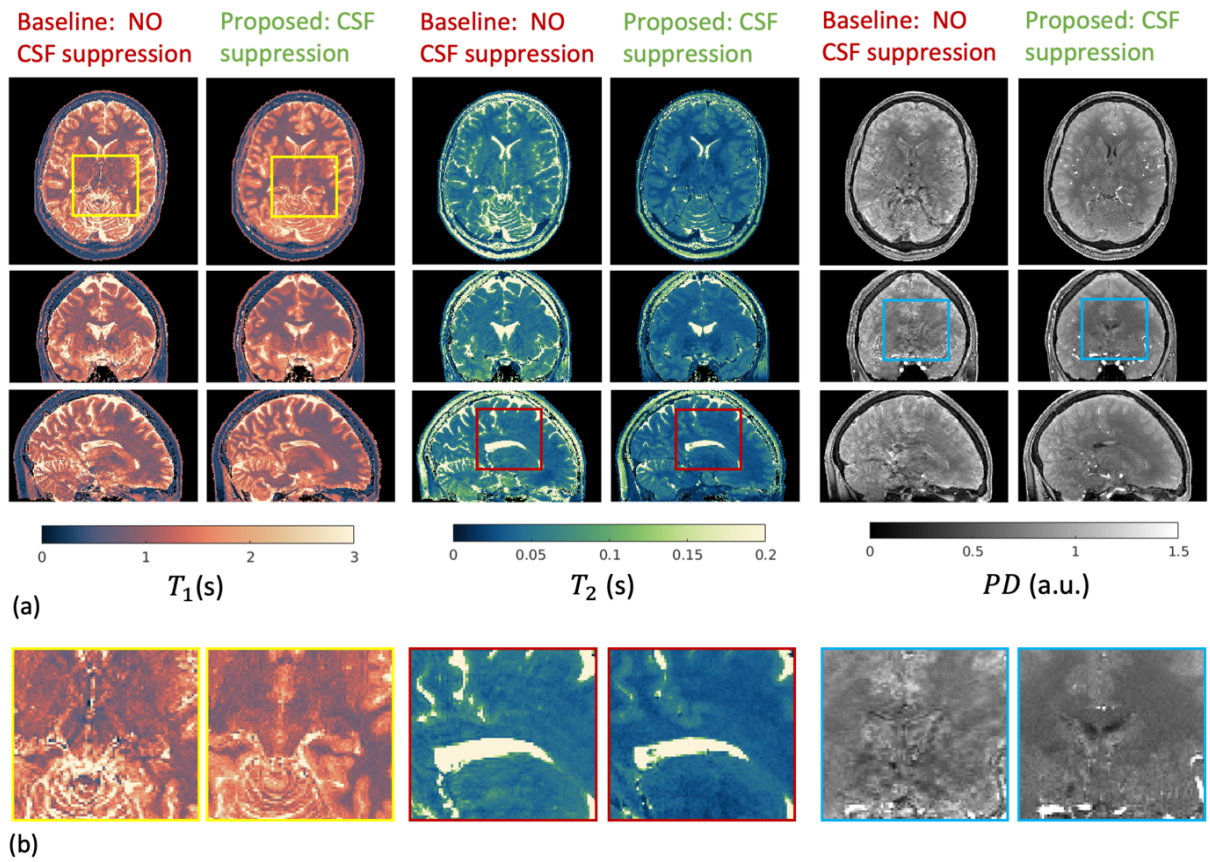

**Figure S3.** In-vivo 3D MR-STAT results using the baseline and the new optimized flip-angle trains. Similar as Figure 3 but for volunteer 3.

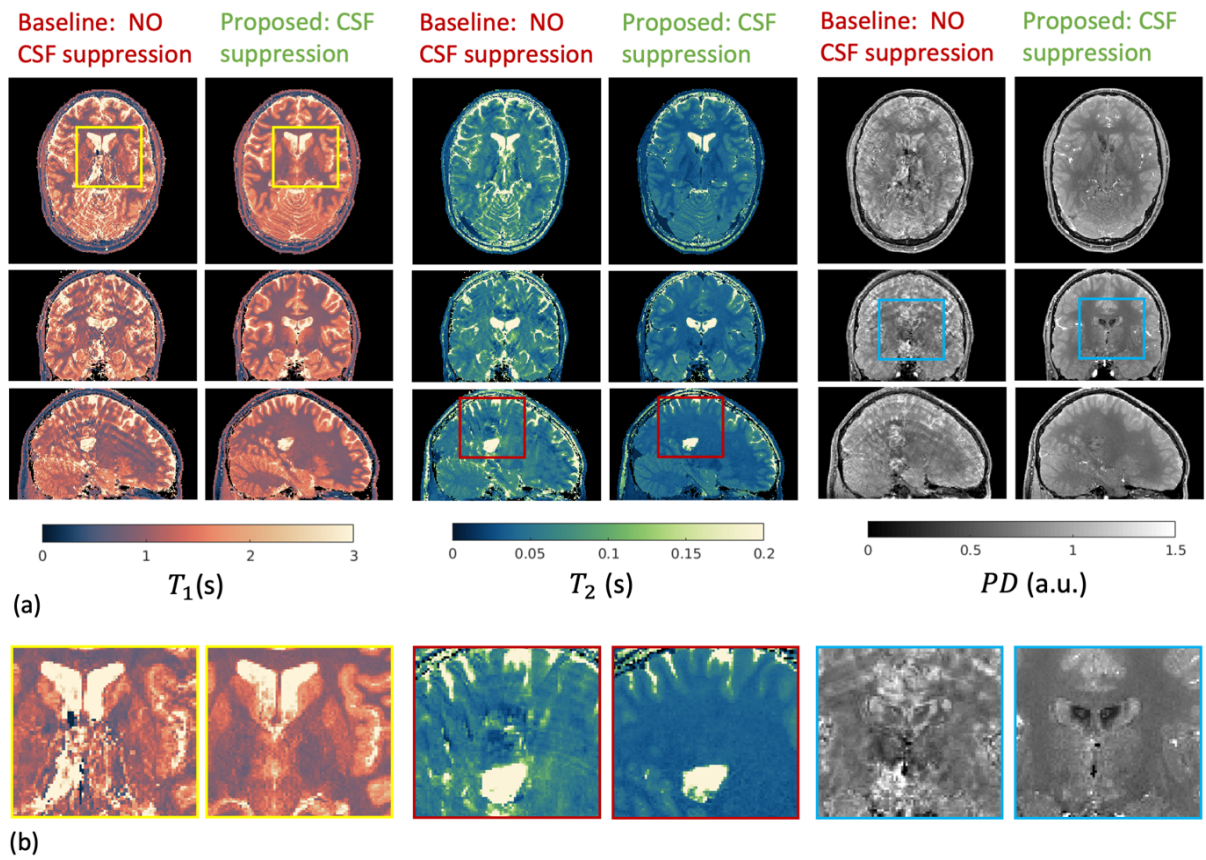

(a)

(b)

**Figure S4.** In-vivo 3D MR-STAT results using the baseline and the new optimized flip-angle trains. Similar as Figure 3 but for volunteer 4.

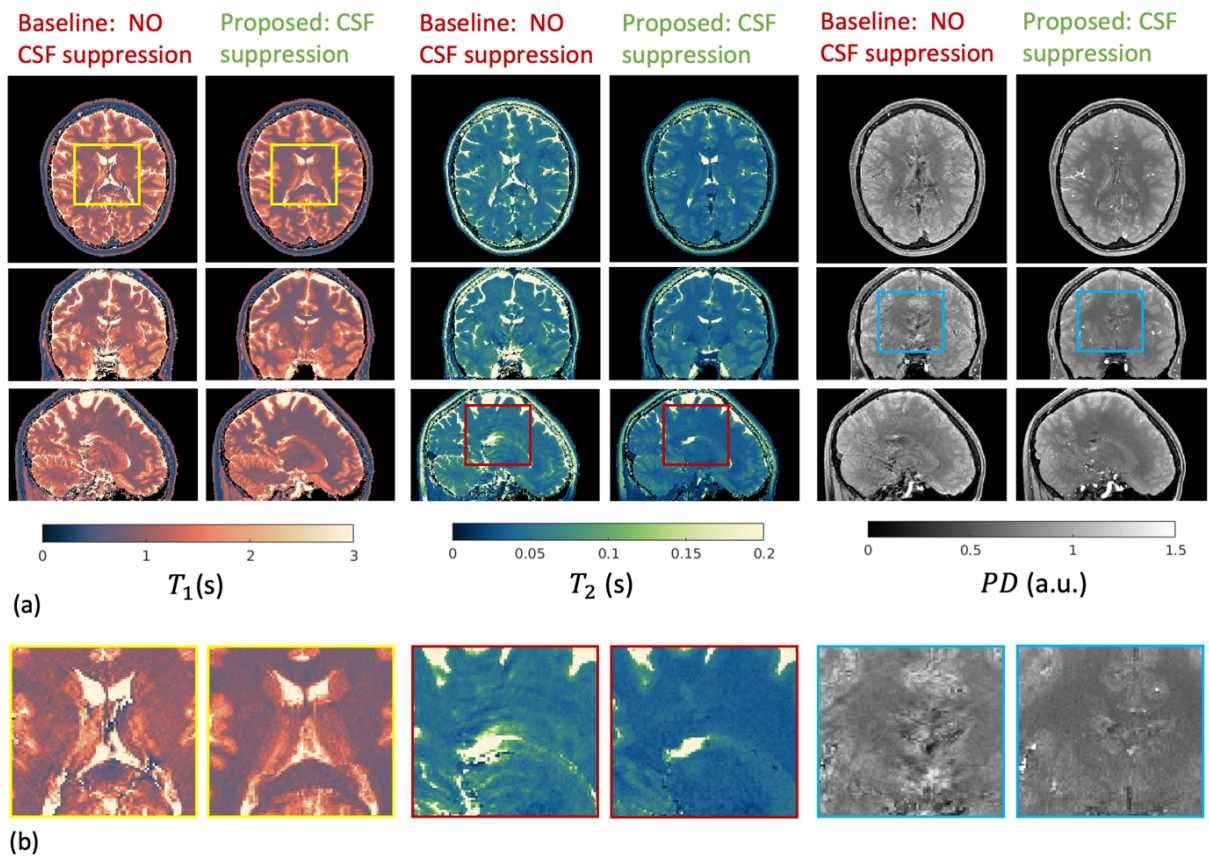

(a)

(b)